\documentclass[11pt,a4paper]{article}
\pdfoutput=1

\usepackage{graphicx}

\usepackage{amsmath}
\usepackage{amssymb}
\usepackage{amsfonts}
\usepackage{graphicx,bm}
\usepackage{dcolumn}

\usepackage{color,amsxtra}
\usepackage{epsf}

\usepackage{enumerate}
\usepackage{hhline}
\usepackage{array}
\usepackage{tabularx}

\usepackage{subfigure}

\usepackage{fancyhdr}
\usepackage{mathrsfs}

\usepackage{amsmath}
\usepackage{amsthm}
\usepackage{amssymb}
\usepackage{amscd}
\usepackage[british]{babel}
\usepackage{babel}
\usepackage{graphicx}
\usepackage{psfrag}
\usepackage{epsfig}
\usepackage{rotating}
\usepackage{times}

\newcommand{\Eqn}[1]{&\hspace{-0.2em}#1\hspace{-0.2em}&}

\theoremstyle{plain}

\newcommand{\boxend}{\flushright{$\Box$}}

\newcommand{\w}{\omega}

\begin{document}

\title{Bouncing  Loop Quantum Cosmology from $F(T)$ gravity}

\author{Jaume Amor\'os $^{1}$\footnote{E-mail: jaume.amoros@upc.edu}, Jaume de Haro$^{1}$\footnote{E-mail: jaime.haro@upc.edu} and Sergei D. Odintsov$^{2,3, 4,  }$\footnote{
E-mail address: odintsov@ieec.uab.es}}

\maketitle

{$^{1}$Departament de Matem\`atica Aplicada I, Universitat
Polit\`ecnica de Catalunya, Diagonal 647, 08028 Barcelona, Spain
\\
$^2$Instituci\'{o} Catalana de Recerca i Estudis Avan\c{c}ats (ICREA),
Barcelona, Spain\\
$^3$Institut de Ci\`encies de l'Espai (CSIC-IEEC),
Campus UAB, Facultat de Ci\`encies, Torre C5-Par-2a pl, E-08193 Bellaterra
(Barcelona), Spain\\
$^4$Tomsk State Pedagogical University, Tomsk, Russia
and Eurasian Nat.University, Astana, Kazakhstan.}

\thispagestyle{empty}

\begin{abstract}
The big bang singularity could be understood as a breakdown of Einstein's General Relativity at very high energies. Adopting this viewpoint,
other theories, that implement Einstein Cosmology at high energies,
might solve the problem of the primeval singularity.
One of them is Loop Quantum Cosmology (LQC)
with a small cosmological constant
that models a universe moving along an ellipse, which prevents singularities like  the big bang or the big rip, in the phase space $(H,\rho)$,
where $H$ is the Hubble parameter and $\rho$ the energy
density of the universe.
 Using LQC when one  considers a model of universe filled by radiation and matter where,
 due to the cosmological constant, there are a de Sitter and an anti de Sitter solution.
This means that one
obtains  a  bouncing non-singular universe which is in the contracting phase at early times. After
leaving this phase, i.e., after bouncing, it passes trough a radiation and matter dominated phase and finally at late times it expands
 in an accelerated way (current cosmic acceleration). This model does not suffer from the  horizon and flatness problems as in big bang cosmology, where
 a period of inflation that increases the size of our universe in more than $60$ e-folds is needed in order
 to solve both problems. The model has two mechanisms to avoid these problems:  The evolution of the universe through a contracting phase 
and a period of super-inflation ($\dot{H}> 0$).
\end{abstract}

{\bf Pacs numbers:}{04.50.Kd, 98.80.-k, 98.80. Jk}


\section{Introduction}
When one considers a universe filled by radiation and matter expanding following the
standard Einstein Cosmology (EC), i.e.
when the  dynamics of the universe is dictated  by the equations of the General Relativity,  coming back in  time,
 one concludes that there exists,
at very early
times, a primeval singularity named big bang. 

The big bang  singularity could be seen as a deficiency of EC at high energies, because there is not any objective reason which
supports the same physics at high than at low energies.
In fact, one can claim that the big bang signals the breakdown of General Relativity at high energy density scales. However,
there are  observational
evidences, such as the discover of the  cosmic microwave background (CMB) by Arno Penzias and Robert Wilson in 1964,  that the  ''big bang model``
 works correctly at scales lower than Planck's. At those scales, the universe is filled by a hot photon-baryon plasma that could be
modelled by a radiation fluid which cools as the universe expands, and non-relativistic matter starts to dominate allowing the formation of
structures.

A possible solution to the big bang singularity could come
from a modification, at high energies, of  Einstein's General Relativity. Since this theory
could be understood as a linear teleparallel theory (recall that  Einstein used teleparallelism in an unsuccessful attempt to unify gravitation with
electromagnetism
\cite{e30}), because
its Lagrangian is a linear function of the spacetime {\it scalar torsion}, namely $T$, one can assume
that our
universe could be described by  non-linear
teleparallel theories ($F(T)$ theories) \cite{Hayashi:1979qx,Hehl:1976kj, Flanagan:2007dc, g10}, that become nearly linear at low energies.

It is known that $F(T)$ gravity can realize both inflation~\cite{ff07}
and the late-time cosmic
acceleration~\cite{bf09, l10, BGL-Comment}, 
revealed by recent observations
for example,
Type Ia Supernovae~\cite{SN1},
baryon acoustic oscillations (BAO)~\cite{Eisenstein:2005su},
large scale structure (LSS)~\cite{LSS},
cosmic microwave background (CMB) radiation~\cite{WMAP},
and effects of weak lensing~\cite{Jain:2003tba} (see \cite{bcno12} for a recent review of current cosmic 
acceleration).
In fact, a very large number of recent papers are devoted to investigate  diverse properties of $F(T)$ gravity
in order to check whether it could be  a veritable alternative to General Relativity \cite{molts}.
 Moreover,
 models of $F(T)$ gravity in which
the finite-time future singularities appear have been
reconstructed ~\cite{bmno12}.

When one considers an homogeneous and isotropic space-time, i.e., when one considers the Friedmann-Lema\^{\i}tre-Robertson-Walker (FLRW) geometry,
 the scalar torsion is given by $T=-6H^2$,
where $H$ is the Hubble parameter \cite{bf09}, as a very remarkable consequence, $F(T)$ cosmologies entail that
the modified Friedmann equation  depicts a curve in the plane $(H,\rho)$,
where $\rho$ denotes the energy density of the universe.
That is, the universe moves along this curve an
its dynamics is given by the so-called {\it modified Raychaudhuri equation} and the conservation equation.
This opens the possibility to build non-singular models of universes with a cosmological constant and  filled by
radiation and matter.
Moreover, $F(T)$ theories could be used to reconstruct cosmologies  in two ways: i) Given the scale factor $a(t)$ and the Equation of State (EoS), one
can build the corresponding   $F(T)$. ii) Given  the scale factor $a(t)$ and the $F(T)$ theory one can build the corresponding EoS.

Our main result is to show that, for the flat FRWL geometry, choosing as $F(T)$ theory  the effective formulation of Loop Quantum Cosmology
(see \cite{lqc} for papers in effective LQC),
the modified Friedmann equation that
includes holonomy corrections  gives, at early times, a universe  in an anti de Sitter phase, which after leaving this phase  starts to
accelerate  leaving the   contracting phase to enter in the expanding one (it bounces),
then it starts to decelerate and passes trough  a radiation and  matter dominated phase. Finally, at late times it enters in a
 de Sitter phase
(late time cosmic acceleration). Our  model does not suffer the flatness and horizon problems
that appear in big bang cosmology, because it has a contracting phase and a super-inflationary period ($\dot{H}>0$), then in principle, making unnecessary an inflationary epoch such as that
of big bang cosmology,
where the scale factor increases more that $60$ e-folds
 in order to solve these problems. Moreover, the
evolution of the universe at early times, in a contracting matter-dominated phase, could produce an scale-invariant spectrum of cosmological perturbations that agrees
whit current observations. Finally it is important to stress that our wievpoint of LQC as a $F(T)$ theory opens the possibility to study perturbations in
LQC using the perturbation equations in $F(T)$ gravity, recently deduced in \cite{Saridakis}. We believe that this fact could be very important because, 
 perturbations with holonomy corrections in LQC were introduced on a phenomenological level by replacing the Ashtekar connection $\gamma \bar{k}$
by $\frac{\sin(m\bar{\mu}\gamma \bar{k})}{m\bar{\mu}}$, $m$ being a number (see for example \cite{Bojowald}). Moreover, to obtain an anomaly-free perturbation 
theory some counter-terms must be introduced in the Hamiltonian constrain \cite{Barrau}, which for vector perturbations give rise to counter-terms
depending on $\bar{k}$, i.e., they are no longer almost periodic function on $\bar{k}$, which seems to be in contradiction with the spirit of LQC.

The paper is organized as follows: In Section II we study EC and we discuss the different ways to deal with the avoidance of the
big bang singularity. Section III is devoted to the study of LQC, showing that its effective formulation gives a bouncing non-singular
model  where the universe evolves from a contracting phase to our current cosmic acceleration. We also show that  this model model does not suffer the
flatness and horizon problems. Finally, in Section IV the reconstruction of cosmologies is considered, in both, via an scalar field and via $F(T)$
theories.

The units  used through the paper are $ c = \hbar = 8\pi G=1$.
\medskip

\section{Einstein cosmology: radiation plus  matter plus cosmological constant}
Assuming that, at large scales, our universe is homogeneous and isotropic leads us to
 consider a flat  FLRW space-time, which metric is given by
\begin{equation}\label{1}ds^2=-dt^2+a^2(t)(dx^2+dy^2+dz^2),\end{equation}
where $a$ is the {\it scale factor}: the quantity that "mesures" the distance between points along time.

For this metric we consider
a universe filled  by radiation plus dust matter, which means that the energy density is given by $\rho=\rho_r+\rho_m$,
where $\rho_r$ is the energy density of the radiation and $\rho_m$ is the energy density of the  matter.  Here, 
as usually we assume that matter is dust (cold dark matter).

For this kind of fluids their  pressures satisfy $P_r=\frac{1}{3}\rho_r$ and $P_m=0$. From the first
principle of thermodynamics or conservation equation $d(\rho V)=-PdV$ ($V=a^3$ being the volume), one obtains the following solutions
 \begin{equation}\label{2}
 \rho_r=\rho_{r,0}{V}^{-4/3} \quad\mbox{and}\quad  \rho_m=\rho_{m,0}{V}^{-1},
 \end{equation}
 where the subindex $0$ means that the quantity is evaluated at the present time, and where we have taken $V_0\equiv 1$.

Now we consider the so-called {\it Benchmark model}, where EC with an small cosmological constant $\Lambda$ is used to
study   our universe  filled by radiation plus matter $\rho=\rho_r+\rho_m$.

Note that, EC can be seen as a linear teleparallel theory with
 Lagrangian \cite{ha13}
\begin{eqnarray}\label{3}
{\mathcal L}_E(T)=\frac{1}{2}TV -(\rho+\Lambda)V,
\end{eqnarray}
where $T=-6H^2$ is the so-called {\it scalar torsion}. Or in its more conventional form
\begin{eqnarray}\label{4}
{\mathcal L}_E(R)=\frac{1}{2}RV -(\rho+\Lambda)V,
\end{eqnarray}
where $R=6(\dot{H}+2H^2)$ is the  {\it scalar curvature}.

In spite that both formulations are equivalent,
it is important to recall that
 teleparallel theories are constructed from the Weitzenb\"ok connection obtaining an space-time with vanishing curvature (the Riemann tensor vanishes)
  but  not torsion free,
in contrast with the standard Levi-Civita connection which gives a curved torsion-free space-time.

From these Lagrangians one easily obtains the Hamiltonian constraint that leads to the
 basic equation in cosmology: The so-called  {\it Friedmann equation}, which in  EC
 is given by
\begin{eqnarray}\label{5}
H^2=(\rho+\Lambda)/3,
\end{eqnarray}
depicting a parabola in the plane $(H,\rho)$, i.e., the evolution of the universe
follows this parabola, and its  dynamics is given by  the system (which could be easily obtained from the conservation and Friedmann equations)
\begin{eqnarray}\label{6}\left\{\begin{array}{ccc}
 \dot{H}&=&-\frac{2\rho_r}{3}-\frac{\rho_m}{2}\\
\dot{\rho}&=&-4H\rho_r-3H\rho_m
,\end{array}\right.
\end{eqnarray}
provided that the universe moves along the parabola $H^2=(\rho+\Lambda)/3$, and that $\rho_r$ and $\rho_m$ satisfy equation   (\ref{2}).
In (\ref{6}) the first equation is the so-called Raychaudhuri equation, and the second one is an equivalent form of the conservation
equation $d(\rho V)=-PdV$.

Equation (\ref{6}) is a first order two-dimensional dynamical system. This kind of  systems have a very simple dynamics that could be easily understood
calculating their
 {\it  critical points} (points in the phase-space $(H,\rho)$ satisfying $\dot{H}=\dot{\rho}=0$), which are
stationary solutions.

 The system (\ref{6}), in the expanding phase ($H>0$), has a unique
 critical point
 $(H=\sqrt{\Lambda/3},\rho=0)$ which is a global attractor (from the second equation of (\ref{6}) one easily deduces that $\rho$ decreases with time). This means that the universe enters in a de Sitter phase
at late times (the late time cosmic acceleration).

On the other hand, at early times the universe is dominated by $\rho_r$ and $\rho_m$. Since we are in the expanding phase $H>0$,
the volume $V$ is an increasing function
of the time. As a consequence, one deduces from (\ref{2}) that at very early times the universe is radiation dominated, and
when the energy density reaches the value $\rho=2\rho^4_{m,0}/\rho^3_{r,0}$  it changes to a matter-dominated phase.

Note also that, since the parabola is an unbounded curve, there is only a critical point
of the system and $\rho$ is a decreasing function. The interesting point is to know if the universe reaches the singularity
($\rho=\infty$) in a finite or infinite time. Solving the system (\ref{6}), using that at early times the
universe is in the radiation dominated phase, gives as a solution
\begin{equation}\label{7}
 H(t)=\frac{H_0}{1+2H_0(t-t_0)}.
\end{equation}

From this solution one concludes that the time from the big bang  to the present is $t_0-t_{big bang}=\frac{1}{2H_0}$.

\vspace{0.5cm}

\begin{figure}[h]
\begin{center}
\includegraphics[scale=0.4]{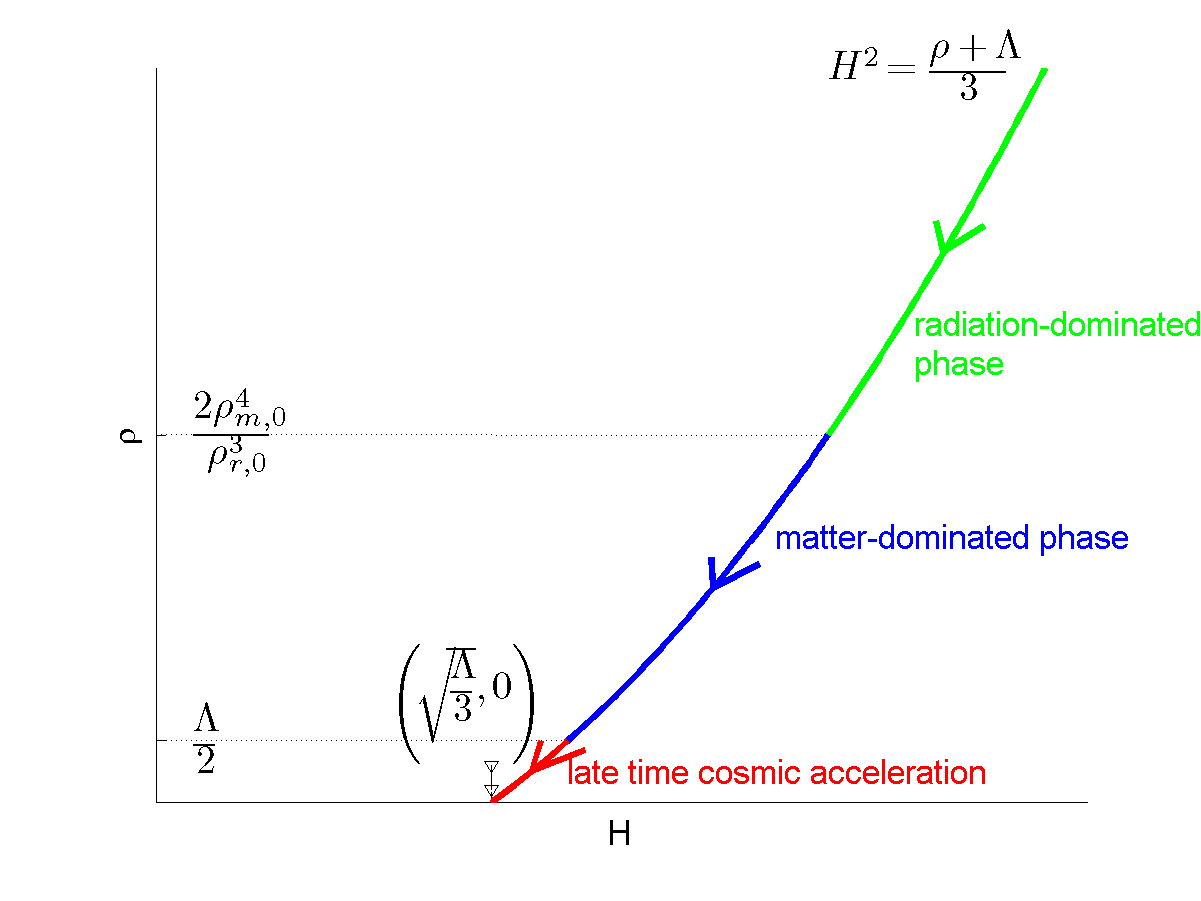}
\end{center}

\caption{{\protect\small The different epochs of the Universe in Einstein Cosmology: a radiation-dominated expansion phase
following the Big Bang, a matter-dominated expansion phase following it, and a final phase describing  the current accelerating expansion.}}
\end{figure}

\subsection{What does exactly the big bang mean?}

At $t_{big bang}$ the energy density diverges ($\rho\rightarrow \infty$). This could be understood as a deficiency of Einstein Cosmology and not
as the beginning of our universe, because Einstein's General Theory of Relativity is, in principle, a  low energy theory. Thus, there is
no objective reason to use this theory at high energies. 

The big bang was discussed towards 1970, when  the idea emerged that quantum effects could be important at
very high energies, leading to
a universe without a primeval singularity \cite{pf73}.  Efforts in this direction gave
rise to the so-called {\it semiclassical gravity}, where quantum effects due to fields coupled with gravity are
taken into account at early times (see for instance \cite{d77}). The most successful model was the so-called "Starobinsky model" \cite{s80}
where the author obtained an unstable nonsingular model in which the universe starts in
the de Sitter phase and ends in a matter dominated phase (the accelerated expansion of the universe
had not been discovered yet at that moment).

Another step in order to deal with the universe at early time was the "inflation theory" \cite{g81}. A beginning of the universe seems incompatible with its homogeneity  and isotropy (the horizon problem), and it is also very
difficult, from a beginning, to understand the present spatial flatness of the universe
(the flatness problem). The underlying idea behind inflation in EC is that, at early times, the universe had  a period where the quantity $aH$ increased considerably.
Since in EC, when the universe is not phantom dominated, $H$ is a decreasing function, to achieve the increase of $aH$ one looks for a mechanism so that
 our universe remains, for a brief period of time, in a quasi de Sitter phase. Then $H$ is nearly constant and the increase of the scale factor is exponential.
The best way to achieve this quasi de Sitter period  is by means of  a field called {\it inflaton},  rolling very slowly according to a potential
at very early times (Planck epoch or later, for example, at
grand unified theories (GUTs) epoch),  producing the   accelerated  expansion of the universe. At the end of this
inflationary
epoch the inflaton field decays creating the matter of the universe which thermalizes,
being the universe in the radiation dominated phase.  Finally, at that epoch, the model
 matches with the standard big bang theory. (It is always said that
the inflationary paradigm is not a theory itself but an implementation to the standard big bang theory). Here, it is important to realize
that the inflationary theory does not deal with
the problem of a initial singularity of our universe because the theory starts at  Planck epoch or, in some models, later.
(Sometimes it is argued that before Planck epoch there is no
classical description of the universe, and only  a quantum description of it is possible).

However, although the inflationary paradigm is the most popular and used by the majority of cosmologists, it  has some problems: i) Inflation
deals with the {\it singularity problem}
in an unconventional way, it effaces all the early history of our universe being itself as a beginning of the universe. In this sense,
one could understand the beginning of the inflation as the beginning of our universe, and it seems impossible to form a previous idea of the
universe before inflation. 
 ii) The {\it amplitude problem} related with the power spectrum of the cosmological
perturbation, as we have seen in chaotic inflation. In a wide class of inflationary models, the potential of the inflation field and the
change of inflation field during inflation, namely $\Delta\phi$, must satisfy the relation
$V(\phi)/(\Delta\phi)^4\leq 10^{-12}$, what imposes a hierarchy in  energy scales. iii) The trans-Planckian problem, that could be formulated as follows:
Inflation provides a mechanism to produce structure formation based on the fact that scales currently observable were originated by wavelengths smaller
than  the Hubble radius at the beginning of inflation. This typically requires that inflation lasts past the scale factor increases $60$ e-folds. However,
if the period of inflation was longer, which happens in the majority of current models, then the wavelengths of all observable scales would be smaller than
the Planck length at the beginning of inflation, but we do not know what kind of physics operates at that scales (see, for instance, \cite{mb01}).


Another completely different way to deal with the initial singularity problem is to assume
EC is only right at low energies, and  then,   in the Lagrangian  (\ref{3})
$T$ has to be changed by  $F(T)$
or in (\ref{4}) $R$ by $F(R)$
where $F$ must be  a nearly linear functions for small values of its argument, to
understand this theory as an implementation of EC at high energies.

The field equations of the teleparallel Lagrangian are of second order, which  is a great advantage compared to the Lagrangian constructed with the scalar
curvature $R$, whose fourth-order
equations lead to pathologies like instabilities or
large corrections to Newton's law \cite{l10}.

This is a good reason to use $F(T)$ teleparallel theories instead the $F(R)$ ones (see \cite{no11} for a recent review of $F(R)$ gravity),
because their simplicity  gives rise to
 modified Friedmann equations   depicting  curves in the plane $(H,\rho)$. According to these theories the universe moves along a curve, and
its dynamics is given by the so-called "modified Raychaudhuri equation" and the conservation equation.

\section{Loop quantum cosmology: radiation plus matter plus cosmological constant}
The standard  viewpoint of LQC  assumes, at quantum level, a discrete nature of space which leads
to a quadratic modification ($\rho^2$)
in its effective Friedmann equation
at high energies \cite{s09a}.
This modified Friedmann equation depicts  the following  ellipse in the plane $(H,\rho)$ (see for details \cite{bho12})
\begin{eqnarray}\label{8}
\frac{H^2}{\rho_c/12}+\frac{(\rho+\Lambda-\frac{\rho_c}{2})^2}{\rho_c^2/4}=1,
\end{eqnarray}
where $\rho_c\equiv \frac{2\sqrt{3}}{\gamma^3}\cong 258.51$
is the so-called critical density, with
$\gamma\cong 0.2375$  being the so-called Barbero-Immirzi parameter \cite{as11}.
Note that, in units used through this paper, Planck's density has the numeric value $\rho_{Planck}=64\pi^2\cong 631.61$ which is greater than $\rho_c$,
and thus, a classical description of the universe seems possible because its energy density  will never exceed  Planck's scale's.

Here an important  remark is in order: Equation (\ref{8}) could be obtained considering the regularized Hamiltonian
\begin{eqnarray}\label{hamiltonian}
H_{LQC} \Eqn{\equiv} -\frac{2V}{\gamma^3\lambda^3}
\sum_{i,j,k}\varepsilon^{ijk}Tr\left[
h_i(\lambda)h_j(\lambda)h_i^{-1} (\lambda) 
h_j^{-1}(\lambda)h_k(\lambda)\{h_k^{-1}(\lambda),V\}\right]+\rho V,
\end{eqnarray}
where $h_j(\lambda)\equiv e^{-i\frac{\lambda \beta}{2}\sigma_j}$ are holonomies being $\lambda=\sqrt{\frac{\sqrt{3}}{4}\gamma}$ 
a paramater with dimensions of length \cite{s09a}, and $\beta$ is the canonically conjugate variable to the volume $V$ satisfying $\{\beta, V\}=
\frac{\gamma}{2}$.

An explicit calculation of this hamiltonian were done in \cite{haro} giving as a result
\begin{eqnarray}\label{hamiltonian}
H_{LQC}= -3V\frac{\sin^2( \lambda \beta)}{\gamma^2\lambda^2}+\rho V.
\end{eqnarray}

Then the Hamilton equation $\dot{V}=\{V,{\mathcal H}_{LQC}\}$ is equivalent to the identity $H= \frac{\sin(2\lambda \beta)}{2\gamma\lambda}$
that, combined with the Hamiltonian constrain $H_{LQC}=0$ give rise to the modified Friedmann equation (\ref{8}) (see for instance, \cite{bho12}). 

\vspace{0.5cm}

The dynamics is now given in LQC by
the system
\begin{eqnarray}\label{9}\left\{\begin{array}{ccc}
 \dot{H}&=&-\frac{4\rho_r+3\rho_m}{6}\left(1-\frac{2(\rho+\Lambda)}{\rho_c} \right)\\
\dot{\rho}&=&-4H\rho_r-3H\rho_m.\end{array}\right.
\end{eqnarray}

In  order to understand the dynamics of the system it is very useful to
introduce the following parameter
 $\omega_{eff}\equiv -1-\frac{2\dot{H}}{3H^2},$
 which in
  LQC becomes
\begin{equation}\label{10}
  \omega_{eff}=-1+\frac{4\rho_r+3\rho_m}{3(\rho+\Lambda)}\frac{\rho_c-2(\rho+\Lambda)}{\rho_c-(\rho+\Lambda)}.
\end{equation}

This quantity
is related to the expansion of the universe. Actually, when $\omega_{eff}<-1/3$
(respectively $\omega_{eff}>-1/3$)
the universe accelerates (respectively decelerates). In fact,  one can see  the universe  filled by an effective fluid
that drives its
 dynamics,   and  whose  pressure and energy density are related by $\w_{eff}=P/\rho$.

Coming back to the system (\ref{9}), note first that at low energies ($\rho\ll \rho_c$) it coincides with the system (\ref{6}), which means that at low energies LQC
coincides with EC, and it could be understood as
an implementation of EC at high energies.
In fact, writing (\ref{9}) in its more usual form
\begin{eqnarray}
H^2=\frac{\rho+\Lambda}{3}\left(1- \frac{\rho+\Lambda}{\rho_c}\right),
\end{eqnarray}
one can see that for the current value of the energy density $\rho_0$, which satisfy $\frac{\rho_0}{\rho_c}\sim 10^{-120}$, one has $H^2=\frac{\rho+\Lambda}{3}$, what means
that nowadays
there is no any visible difference with standard $\Lambda$CDM cosmology.

Studying (\ref{9}) as a dynamical system we can see that it
 has two critical points
$p_f\equiv (\sqrt{\frac{\Lambda}{{3}}}\sqrt{\left(1-\frac{\Lambda}{\rho_c}\right)},0)$ and
$p_i\equiv(-\sqrt{\frac{\Lambda}{{3}}}\sqrt{\left(1-\frac{\Lambda}{\rho_c}\right)},0)$. The first one is a de Sitter solution and the second one is
 an anti de Sitter solution.
The universe moves along the ellipse from $p_i$ to $p_f$ in a clockwise sense (this comes from the second equation of (\ref{9}), because in
the contracting phase the energy density is an increasing function and in the expanding one it is decreasing).
At very early times the size of the universe was very large and it contracts with positive acceleration because for $\rho\sim 0$ one has
$\omega_{eff}\sim -1<-1/3$. When the cosmological constant $\Lambda$ stops its domination, the universe enters in a contracting matter dominated phase ($\omega_{eff}\sim 0$)
 because the volume
is still big enough. Then the volume decreases and the universe enters in the contracting radiation dominated phase
($\omega_{eff}\sim 1/3$). In the contracting phase, as we have already showed, $\rho$ is an increasing function and  when $\rho\sim \rho_c/3$ one has
$\omega_{eff}\sim -1/3$ which means that the universe accelerates (that is, it contracts in a decelerating way). This behavior is due to the
 form of the ellipse and it could be
understood as a sort of dark energy that drives  our universe to this accelerated phase.
In this phase,
when it arrives at the point $p_1=(\rho_c-\Lambda,0)$
(the top of the ellipse),
 it bounces  leaving the contracting phase and entering  in the
expanding one where the energy density starts to decrease.
At that moment one has $\omega_{eff}\ll -1/3$ and thus the universe expands in an accelerating way, it is in a {\it super-inflationary phase}
that  only increases the size of the universe by a small number of e-folds, which is not enough to solve the flatness and horizon
problems that appear in EC. However, as we will show in next Section, our model does not suffer from these problems.
This accelerating period finishes when the universe  arrives at $p_2\cong (\rho_c/3,\rho_c/3)$. At that moment,  it starts to
decelerate and when the density satisfies $\rho_c\ll \rho\ll \Lambda$ the universe enters first in a radiation
dominated phase, which it leaves when $\rho=2\rho^4_{m,0}/\rho^3_{r,0}$, to enter in a matter dominated one
 ($\omega_{eff}\cong 0$). Finally, after leaving this phase, it enters in an accelerated phase
when $\rho=\Lambda/2$ ($\omega_{eff}<-1/3$) and goes asymptotically, at late times, to the point $p_f$  (de Sitter phase that mimics the late time
accelerated cosmic expansion).

\vspace{0.5cm}

\begin{figure}[h]
\begin{center}
\includegraphics[scale=0.6]{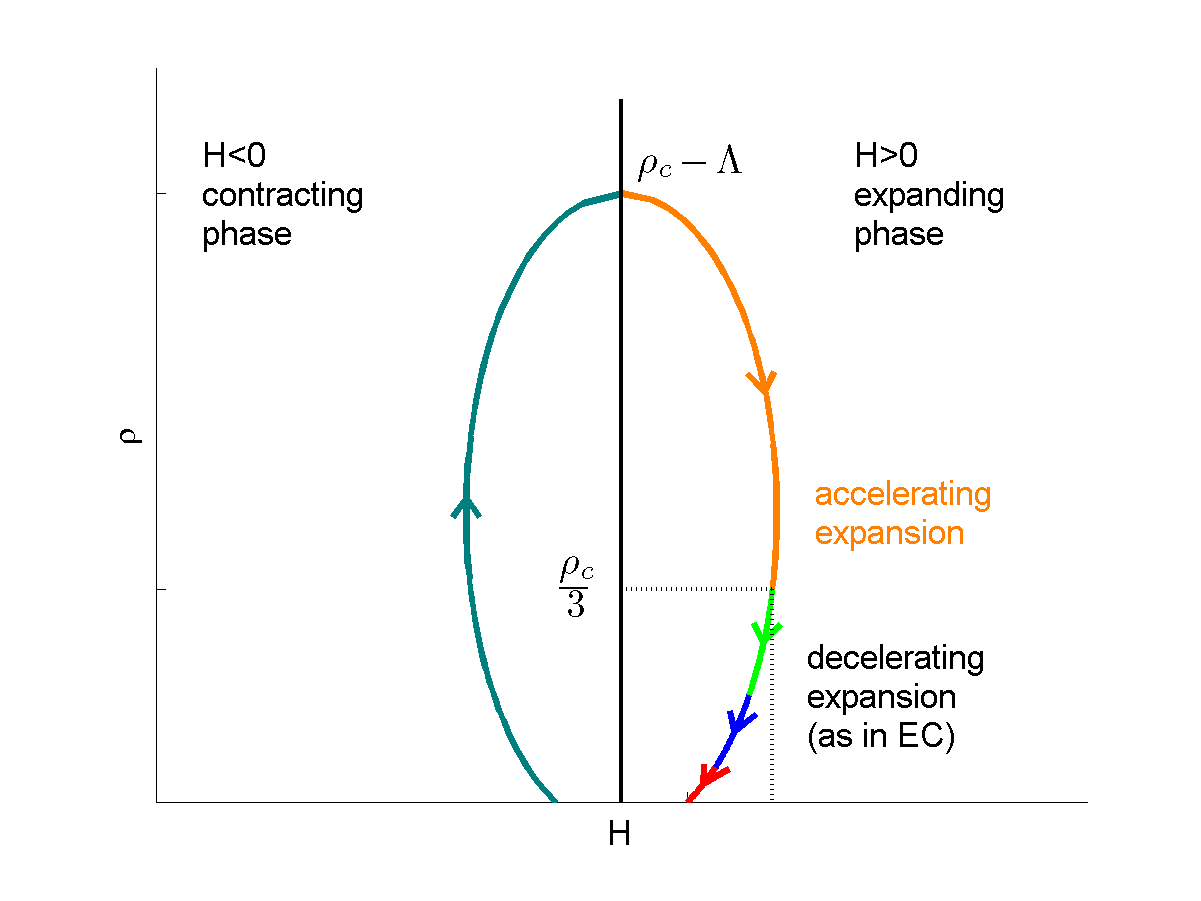}
\end{center}

\caption{{\protect\small The different epochs of the Universe in Loop Quantum Cosmology: after a contracting phase and
an accelerated expansion phase, the Universe enters a decelerating expansion phase as in Einstein Cosmology.}}
\end{figure}

\subsection{Does this model need an inflationary epoch as in EC?}
In our model the inflationary epoch in the expanding phase starts at bouncing time, namely $t_i$, when the universe has energy density
$\rho_c-\Lambda\cong \rho_c$. At that moment the scale factor
 is minimum, and thus, we can assume that the universe is radiation dominated. As we have seen the universe stops accelerating when the energy density is approximately equal to
$\rho_c/3$. Let $t_f$ be the time when inflation ends, then from (\ref{2}) one deduces $a(t_f)=3^{1/4}a(t_i)$. But in EC, to solve the
horizon and flatness problem one needs and amount in the scale factor greater than $60$ e-folds \cite{g81}, what clearly does not happen in our model.

In fact,
for a fluid with linear EoS $P=\omega \rho$ where $\omega>-2/3$, the accelerated expansion ends when $\rho\cong \frac{\rho_c(1+3\omega)}{2(2+3\omega)}$.
Then, a simple calculation yields
\begin{eqnarray}
a(t_f)=\left(\frac{2(2+3\omega)}{1+3\omega}\right)^{\frac{1}{3(1+\omega)}}a(t_i),
\end{eqnarray}
which means that to obtain $60$ e-folds one needs a value of $\omega$  very close to $-2/3$. Note also that for fluids with $\omega>0$ one obtains a "{\it bad inflation}" (an inflation
with an small increase of the scale factor).

However, it is important to realize that in our model these problems do not appear.
i) First, we start with the horizon problem. To simplify the calculation we assume a matter dominated universe. The particle horizon
in the contracting phase is
\begin{eqnarray}
d_{hor}=a(t_c)\int_{-\infty}^{t_{c}}\frac{dt}{a(t)},
\end{eqnarray}
where $t_c$ is the bouncing time.

Using the identity $\rho(t)V(t)=\rho(t_c)V(t_c)$ and the conservation equation $\dot{\rho}=-3H\rho$ one obtains
\begin{eqnarray}
d_{hor}=\frac{\rho_c}{\rho^{1/3}(t_c)}\int_{0}^{\rho_{c}-\Lambda}\frac{d\rho}{(\rho+\Lambda)\rho^{2/3}(\rho_c-(\rho+\Lambda)))}\nonumber\\
\sim \frac{1}{\rho_c}\int^{\rho_{c}-\Lambda}\frac{d\rho}{(\rho_c-(\rho+\Lambda)))}=+\infty.
\end{eqnarray}
which means that, when the universe enters in the expanding phase, all the points of it are in causal contact and, thus,
the universe has had enough time to be homogeneous and isotropic
when it bounces. Note that, the same result was deduced in \cite{ff07} where the authors studies the teleparallel version of
the Born-Infeld Lagrangian,
${\mathcal L}_{BI}=\frac{1}{2}V\lambda\left[\sqrt{1+\frac{2R}{\lambda}}-1\right]$, with $\lambda$ being a parameter introduced with the aim of
smoothing singularities. ii)
The flatness problem in EC goes as follows: For a spatially curved  FLRW  space-time the Friedmann equation, in EC,  can be written as
\begin{eqnarray}
\Omega-1=\frac{1}{\dot{a}^2}=\frac{1}{a^2H^2},
\end{eqnarray}
where $\Omega=\frac{\rho+\Lambda}{3H^2}$. In EC cosmology $\dot{a}^2$ is a decreasing function because $\frac{d}{dt}\dot{a}^2=2\ddot{a}\dot{a}<0$. Since
nowadays one has $|\Omega-1|\leq 0.2$ one
easily deduces that at Planck scales $|\Omega-1|\sim 10^{-60}$. From this result it seems that it would be far better to find a physical mechanism
for flattering the universe,
instead of relying on contrived initial conditions at Planck epoch. In EC this problem is solved with a brief period of inflation ($\ddot{a}>0$)
after Planck's epoch. If the number of
e-folds is large enough, then assuming that  $|\Omega-1|\sim 1$ at Planck's epoch one obtains, for the majority of current inflationary models,
$|\Omega-1|\ll  10^{-60}$. However,
our model contains its own  mechanism to solve that problem. Namely, we consider in order to simplify  a matter dominated universe without
cosmological constant (although our reasoning is completely general). Then the solution of the system
(\ref{9}) is given by \cite{h12}
\begin{eqnarray}
H(t)=\frac{\rho_ct/2}{3\rho_ct^2/4+1},\quad \rho(t)=\frac{\rho_c}{3\rho_ct^2/4+1},
\end{eqnarray}
where here we have chosen as a bouncing time $t=0$. From these values one easily find the following scale factor
$a(t)=a(0)\left(\rho(t)/\rho_c\right)^{-1/3}$.

Near the bouncing time $t\sim 0$,
thus $a(t)\cong a(0)$ and $H(t)\cong \rho_c t/2$ and consequently
\begin{eqnarray}
\Omega-1\cong \frac{4}{a^2(0)\rho_c^2t^2}\gg 1,
\end{eqnarray}
that is, the fine tuning of $\Omega-1$ is not needed at any scale.

As one can easily see,
this situation is very different from inflation in EC. Since in EC  $H$ decreases for non-phantom universes, one needs  a brief period of time where the Hubble parameter is nearly constant and the scale factor sustains a huge increase.
In LQC, at high energies the universe is in a super-inflationary phase ($\dot{H}>0$). Then to solve the flatness (and also the horizon) problems one only needs a huge increase of
$aH$. In fact, to solve these problems one needs that $\bar{N}\equiv \ln\frac{a(t_f)H(t_f)}{a(t_i)H(t_i)}\sim 60$,  where $t_i$ and $t_f$ are, respectively, the beginning and end of the
inflationary period \cite{lpb94,cmns08}. And, since in LQC  $H\cong 0$ near the bounce, one always obtains $\bar{N}\gg 1$. Finally, note that if inflation was
produced in a quasi de Sitter phase,  $\bar{N}$ will coincide
with the standard quantity that measures the number of e-folds in inflationary EC. I.e., $\bar{N}$ will coincide with ${N}\equiv \ln\frac{a(t_f)}{a(t_i)}$.

Dealing with the problem of the origin of density perturbations is a different subject.
 One can assume initial conditions, at very early times, for the density perturbations
and one shows that, at late times, they evolve into a scale-invariant spectrum, or
one has to
look for a mechanism that produces an almost scale-invariant spectrum of cosmological  perturbations.
  In this second case, one may consider a condensate scalar field (the inflaton field), and use its quantum fluctuations 
 at high energy
 scales in order to explain the generation of large-scale perturbations. The fact that $H(t)$ is almost constant during the  slow-roll period 
 means that it is possible to generate
 scale-invariant density perturbations on large scales.

The alternative possiblitiy we propose is to consider initial perturbations, for example given by
quantum fluctuations due to a very light field minimally coupled with gravity
  (the quantum fluctuations of the inflaton and the quantum fluctuations of a massless minimally coupled  field  satisfy
 the same Klein-Gordon equation), in our model.

Since in a contracting, matter-dominated phase  of a bouncing universe, cosmological perturbations, have been studied in the last decade,
showing analytically and numerically in some toy models that
they evolve into a scale-invariant
spectrum of cosmological perturbations at late times (after the bounce) \cite{w99,b09, pp08,pp02}. 
In our model we can consider, at very early times,
 quantum fluctuations
that  at  the contracting matter dominated phase  would produce on long wavelengths (at scales larger than the Hubble radius)
an scale-invariant spectrum which would survive after the bounce. This is, of course, a topic that needs future detailed investigation,
 but in principle, from previous works, it seems plausible that our model provides a scale-invariant spectrum after the bounce.

All these reasons indicate that models such as non-singular bouncing cosmologies, where inflation is not needed, should  be taken into account
in order to explain the evolution of our universe.

\section{Reconstructing  cosmologies}
In this section we take another viewpoint: Given the evolution of our universe, i.e., choosing  the evolution of scale factor,
we will  construct the Lagrangian whose dynamical equations have as a solution the chosen scale factor.

\subsection{Reconstruction via an scalar field}
First at all, we consider, in EC, a scalar field $\phi$ with energy density and pressure given by
\begin{eqnarray}\label{17}
\rho=\frac{1}{2}\omega(\phi)\dot{\phi}^2+V(\phi),\quad P=\frac{1}{2}\omega(\phi)\dot{\phi}^2-V(\phi),
\end{eqnarray}
where $\omega$ and $V$ are functions of the field $\phi$.
After some algebra one obtains the relations
\begin{eqnarray}\label{18}
\omega(\phi)\dot{\phi}^2=-2\dot{H},\quad V(\phi)=3H^2+\dot{H}.
\end{eqnarray}

Equation (\ref{18}) has two different solutions. i) If one takes $\omega(\phi)\equiv 1$ then one has \cite{p02}
\begin{eqnarray}
 V(t)=3H^2+\dot{H},\quad {\phi}(t)=\int dt\sqrt{-2\dot{H}}.
\end{eqnarray}

These equations determine $\phi(t)$ and $V(t)$ in terms the scale factor, thereby implicitly determining $V(\phi)$. ii) Taking $\phi=t$ \cite{no06}, which gives
\begin{eqnarray}\label{20}
 V(t)=3H^2+\dot{H},\quad \omega(t)=-2\dot{H},
\end{eqnarray}
once again, any cosmology with scale factor $a(t)$ is realized by the potential $V$.

As an example,
a power law expansion $a(t)=a_0\left|t/t_0\right|^p$  is obtained using formulas (\ref{18}) from an exponential
potential of the form
\begin{eqnarray}
V(\phi)=e^{-\sqrt{\frac{2}{p}}\phi}.
\end{eqnarray}

However,  realistic cosmologies require very complicated potentials that in general do not have a minimum as that of potentials used in inflation.
Then, in general, the scalar field does not oscillate around the minimum and consequently does not release its energy producing light particles that thermalize our universe as occurs in inflationary cosmologies.  In order to obtain a realistic re-heating
theory, one has to use  gravitational particle production. Gravitational particle production due to a transition from a de Sitter to a radiation phase has been studied extensively in the past. Given a consistent re-heating temperature \cite{dv96,s93,pv99}, then  it seems mandatory that, reconstructing models via a scalar field, this transition occurs.

Different examples  reconstructing the history of our universe are given in \cite{enosf08}. Here we study one of them in order to show the complicated potentials obtained:

The dynamics $H(t)=\frac{H_i+\lambda e^{\alpha t}}{1+e^{\alpha t}}$, where $\lambda$, $H_i$ and $\alpha$    are constants satisfying $\alpha, \lambda\ll H_i$
so that slow-roll conditions can be satisfied,
describes a universe which at early times is dominated by an effective cosmological constant with value $3H^2_i$ driven inflation, and
at late times is dominated by another cosmological constant with value $3\lambda^2$ given the current accelerated expansion of our universe. Then, using (\ref{20}) one obtains the following complicated quantities
\begin{eqnarray}&&
\omega(\phi)=\frac{\alpha(H_i-\lambda)e^{\alpha \phi}}{(1+e^{\alpha \phi})^2},\nonumber \\
V(\phi)&=&\frac{3H_i^2+[6H_i\lambda-\alpha(H_i-\lambda)]e^{\alpha \phi}+\lambda^2e^{2\alpha \phi}}{(1+e^{\alpha \phi})^2}.
\end{eqnarray}

\subsection{Reconstruction via $f(T)$ gravity}
In a flat FLRW space-time  filled by a perfect fluid with energy density $\rho$, general teleparallel theories
are obtained from
the Lagrangian
\begin{equation}\label{21}{\mathcal L}=VF(T)-V\rho.
\end{equation}

The conjugate momentum is  then given by
$p_V=\frac{\partial {\mathcal L}}{\partial\dot{V}}= -4HF'(T)$, and thus the Hamiltonian is
\begin{eqnarray}\label{22}
{\mathcal H}=
\dot{V}p_V- {\mathcal L}= (2TF'(T)-F(T)  +\rho )V.\end{eqnarray}

In general relativity  the Hamiltonian is constrained to be zero,
what leads to the modified
Friedmann equation
\begin{eqnarray}\label{23}
\rho=-2 F'(T)T+F(T)\equiv G(T),
\end{eqnarray}
which is a curve in the plane $(H,\rho)$

Then, given a curve of the form $\rho=G(T)$
for some function $G$, a first way to reconstruct the Lagrangian (\ref{21}) consists in
 integrating    the modified Friedmann
equation  (\ref{23}), obtaining as a result
\begin{eqnarray}\label{24}
F(T)=-\frac{\sqrt{-T}}{2}\int \frac{G(T)}{T\sqrt{-T}}dT.
\end{eqnarray}

The simplest example is to take as a curve a parabola, for example,
\begin{eqnarray}
\rho=\bar{\rho}\left(1-\frac{3H^2}{\Lambda}\right),
\end{eqnarray}
which models for a non-phantom universe, i.e., for $\frac{P}{\rho}\geq-1$, a universe that moves clockwise from $(-\sqrt{\Lambda/3},0)$ to $(\sqrt{\Lambda/3},0)$, bouncing when
$(0,\bar{\rho})$.
Using the formula (\ref{24}) one obtains
\begin{eqnarray}
F(T)=\bar{\rho}\left(1-\frac{T}{2\Lambda}\right).
\end{eqnarray}
In this case, if one considers a matter dominated universe and inserts in the conservation equation $\dot{\rho}=-3H\rho$
the value of $H$ as a function of $\rho$, one obtains a solvable differential equation whose solution is
\begin{eqnarray}
\rho(t)=\bar{\rho}\frac{4e^{-\sqrt{3\Lambda t^2}}}{(1+e^{-\sqrt{3\Lambda t^2}})^2},\quad H_{\pm}(t)=\pm\sqrt{\frac{\Lambda}{3}}
\frac{1-e^{-\sqrt{3\Lambda t^2}}}{1+e^{-\sqrt{3\Lambda t^2}}},
\end{eqnarray}
where we have chosen as a bouncing time $t=0$.

As a second example we consider LQC, where
the curve (\ref{8}) can be written in two pieces $\rho=G_-(T)$ (which corresponds to energy densities below $\rho_c/2-\Lambda$) and $\rho=G_+(T)$
(which corresponds to energy densities between $\rho_c/2-\Lambda$ and $\rho_c-\Lambda$), where
\begin{eqnarray}
G_{\pm}(T)=-\Lambda+\frac{\rho_c}{2}\left(1\pm\sqrt{1+\frac{2T}{\rho_c}}  \right).\end{eqnarray}

 Then, using  formula (\ref{24})
 one
gets
\begin{eqnarray}
 F_{\pm}(T)=\mp\sqrt{-\frac{T\rho_c}{2}}\arcsin\left(\sqrt{-\frac{2T}{\rho_c}}\right)
+\frac{\rho_c}{2}\left(1\pm\sqrt{1+\frac{2T}{\rho_c}}
\right)-\Lambda.
\end{eqnarray}

From this formula one obtains, in LQC, the Lagrangian that models a universe with cosmological constant filled by radiation and matter
\begin{eqnarray}
 {\mathcal L}(V,\dot{V})=\left\{\begin{array}{ccc}
   F_{-}(T)V-\rho_{r,0}V^{-1/3}-\rho_{m,0} &\mbox{for} &  0\leq \rho_{r,0}V^{-4/3}+\rho_{m,0}V^{-1}\leq \rho_c/2-\Lambda   \\
    & &  \\  F_{+}(T)V-\rho_{r,0}V^{-1/3}-\rho_{m,0} &\mbox{for} &
 \rho_c/2-\Lambda < \rho_{r,0}V^{-4/3}+\rho_{m,0}V^{-1}\leq \rho_c-\Lambda,
\end{array} \right.
\end{eqnarray}
which shows that the effective formulation of LQC can be considered as a teleparallel theory.

Coming back to
 Formula (\ref{24}), it seems very useful to construct simple bouncing models. One only has to consider a closed curve
in the phase-space $(H,\rho)$. This curve has to be symmetric
with respect to the axe $H = 0$. Splitting the curve in some points as
we have done in LQC, one will easily obtains a $F(T)$ theory
for each part of the curve.

A second way to reconstruct a model using $f(T)$ theories is as follows: given the scale factor $a(t)$, the conservation equation $d(\rho V)=-Pd(V)$ and
the Equation of State $P=P(\rho)$, one obtains the energy density as a function of time $\rho(t)$. From the scale factor $a(t)$ one also obtains the
scalar torsion as a function of time $T=T(t)=-6\left(\dot{a}(t)/a(t)\right)^2$. Then performing the change of variable $T=T(t)$ in (\ref{24}) one obtains
\begin{eqnarray}\label{27}
F(T)=-\frac{\sqrt{-T}}{2}\int^{t(T)} \frac{\rho(s)\dot{T}(s)}{T(s)\sqrt{-T(s)}}ds,
\end{eqnarray}
where  the time $t$ as a function of $T$, i.e. $t(T)$,  was to be obtained inverting the equation $T=T(t)$.

Finally, note that as in the case of an scalar field, formula (\ref{27}) shows that realistic cosmologies, i.e.  realistic $a(t)$  will require of
 a very complicated
$f(T)$ theory.

The last way to construct models has recently been introduced in \cite{bho12}. The idea is that given a scale factor $a(t)$, from the modified Friedmann and Raychaudhuri
equations of a $F(T)$ theory, one can build the corresponding equation of state (EoS) that we will assume has the form $P(\rho)=-\rho-f(\rho)$. To be precise,
 taking the derivative with respect to time in
(\ref{23}) and using the conservation equation one obtains the Raychaudhuri equation
\begin{equation}\label{28}
 \dot{H}=-\frac{f(\rho)}{4}(G^{-1})'(\rho).
\end{equation}

Then, from (\ref{23}) one obtains the time $t(\rho)$ as a function of the energy density. Inserting this expression in $(\ref{28})$ one finally
obtains $f(\rho)$, and thus, one has built the EoS that gives the dynamics $a(t)$ in the corresponding $F(T)$ theory.

As an example we consider in EC ($F(T)=T/2$),  the dynamics
\begin{equation}\label{29}
 H(t)=H_i+H_1e^{-\gamma (t-t_i)}, \mbox{ for }  t_i\leq t\leq 60 H_i^{-1}+t_i,
\end{equation}
where we assume $H_1\ll H_i$ and $\gamma H_i^{-1}\ll 1/60$, which means that $H(t)$ is nearly constant during this period of time, and consequently the
scale factor increases the required $60$ e-folds to solve the horizon and flatness problems.

From (\ref{23}) and (\ref{28}) one easily obtains the following  nonlinear EoS
\begin{equation}
 f(\rho)=2\gamma H_i\left(1-\sqrt{\rho/(3H_i^2)}\right),
\end{equation}
when $\rho\in \left[3(H_i+H_1)^2, 3(H_i+H_1e^{-60\gamma H_i^{-1}} )^2\right]$.

This opens the possibility to consider models where the EoS is nonlinear. One of these models was studied in \cite{ha13}, where in EC with an small cosmological constant $\Lambda$, the
following EoS was considered: $f(\rho)=-\rho\left(1-\rho/\rho_i\right)$. In this case the point $(\sqrt{(\rho_i+\Lambda)/3}, \rho_i)$ is a de Sitter solution, and
the universe evolves from it, passing through a matter-dominated phase,  to  the point $(\sqrt{\Lambda/3},0)$ which mimics the late time cosmic acceleration.
In this case
$\w_{eff}=P(\rho)/\rho=-\rho/\rho_i$, which means that the universe accelerates when $\rho\in [\rho_i/3,\rho_i]$ and decelerates when $\rho \in [0,\rho_i/3]$.
Finally, note that  this model does not contain the horizon and flatness problems. The first one is avoided because at the end of the accelerating phase
all the points of
the universe are in causal contact ($d_{hor}=+\infty$), and the second one due to the accelerated period that reduces the value of $|\Omega-1|$ at
early times.

To finish, we consider once again the dynamics $H(t)=\frac{H_i+\lambda e^{\alpha t}}{1+e^{\alpha t}}$ in EC, and we try to find the EoS.
From the Friedmann equation one obtains
\begin{eqnarray}
e^{\alpha t}=\frac{\frac{H_i}{\lambda}-\frac{\rho}{3\lambda^2}+\sqrt{\frac{\rho}{3\lambda^2}}\left(\frac{H_i}{\lambda}-1\right)}
{\frac{\rho}{3\lambda^2}-1}.
\end{eqnarray}

Then, inserting this value in the Raychaudhuri equation $\dot{H}=\frac{f(\rho)}{2}$ one obtains the function $f(\rho)$. The calculation is easy but
cumbersome, and the final result is a  non-linear EoS given by
\begin{eqnarray}
P(\rho)=-\rho+2\alpha\left({\sqrt{\frac{\rho}{3}}-\lambda}\right)\frac{H_i-\sqrt{\frac{\rho}{3}}}{H_i-\lambda}, \quad \mbox{for}\quad
3\lambda^2\leq \rho\leq 3H_i^2.
\end{eqnarray}

\section{Conclusions}

A large number of models describing non-singular universes could be constructed in $F(T)$ gravity. In this paper we have chosen
LQC (a $F(T)$ theory as we have already showed) with an small cosmological constant
to propose a non-singular bouncing universe filled by radiation and matter, which at late times mimics the current cosmic acceleration. Our model does not suffer the horizon and flatness problems,
so it does not need a quasi de Sitter phase producing a huge increase in the scale factor as it must happen in
EC. Moreover, since at early times our model passes through a contracting matter-dominated phase it could, although this is a complicated point that deserves future investigations, be possible to generate an scale-invariant spectrum
of perturbations.

The development of LQC as a $F(T)$ theory allows the study of LQC perturbations using 
the perturbation equations in $F(T)$ gravity. This is an alternative to the study of
perturbations in LQC up to the present, which is based on phenomenological 
corrections. The authors will pursue this topic in a subsequent work.

However,
teleparallel theories are based in an arbitrary choice of an orthonormal basis, namely $\{{\bf e}_{j}: {j}=0,1,2,3\}$,
in each point of the space-time. For example the particular choice
of the basis $\{{\bf e}_0=\partial_{t}, {\bf e}_1=a^{-1}(t)\partial_{x}, {\bf e}_2=a^{-1}(t)\partial_{y}, {\bf e}_3=a^{-1}(t)\partial_{z}\}$, where  $\partial_t,\dots,\partial_z$ are the vectors corresponding to the cartesian axis in coordinates $(t,x,y,z)$,
gives as a result the scalar torsion $T=-6H^2$, but other different choices (local choices) give another different scalar torsion
\cite{lsb11}, and thus, other completely different cosmologies. 

Fortunately, cosmology based in $F(T)$ gravity does not need that election. Effectively, in cosmology one assumes, at large scales, an homogeneous space-time, which means that the
basis   $\{{\bf e}_j:j=0,1,2,3\}$ could only have a time dependence, because the scalar torsion must be only a function of the time. As a consequence, all admissible bases are related
by time-dependent Lorentz transformations, i.e., by transformations of the form $\Lambda_j^k(t)$, and for these admissible basis it is easy to show that $T$ is invariant with the
value $T=-6H^2$.

\vspace{0.25cm}
{\bf Acknowledgments}
The authors want to thank M. Bojowald and E.N. Saridakis for correspondance and useful comments about cosmological perturbations.
This investigation has been
supported in part by MINECO (Spain), project MTM2011-27739-C04-01, MTM2012-38122-C03-01 and FIS2010-15640,
 and by AGAUR (Generalitat de Ca\-ta\-lu\-nya),
contracts 2009SGR 345, 994 and 1284.



\end{document}